# MOORE'S PARADOX AND THE LOGIC OF BELIEF

**Andrés Páez**
Department of Philosophy
Universidad de los Andes

ABSTRACT

Moore's Paradox is a test case for any formal theory of belief. In *Knowledge and Belief*, Hintikka developed a multimodal logic for statements that express sentences containing the epistemic notions of knowledge and belief. His account purports to offer an explanation of the paradox. In this paper I argue that Hintikka's interpretation of one of the doxastic operators is philosophically problematic and leads to an unnecessarily strong logical system. I offer a weaker alternative that captures in a more accurate way our logical intuitions about the notion of belief without sacrificing the possibility of providing an explanation for problematic cases such as Moore's Paradox.

## 1. Introduction

Among the problems that any logical analysis of the notion of *belief* has to address, Moore's Paradox occupies a preeminent position. G. E. Moore observed that sentences such as 'It's raining but I don't believe it' cannot be used to make coherent assertions, even though they are not actual contradictions (1942, pp. 541-543; 1944, p. 204). There are situations in which the sentence will be true but none in which anybody could use it in a literal sense. In general, sentences of the form:

(1)     '*p* but I do not believe that *p*'

are not self-contradictory, but there are no circumstances in which one can use them to perform coherent assertoric speech acts. The divergence between the truth conditions and the performance conditions of (1) leads to the paradoxical result that there are true sentences that one cannot utter.[1]

---

[1] Moore's Paradox can be extended to other propositional attitudes (Searle 1969). For example, the statement 'Rain is likely, but I do not expect it' is also absurd. In this essay I will only be concerned with the notion of belief.

That a sentence of this type is not self-contradictory is illustrated by the fact that a simple change of person turns (1) into the perfectly natural sentence

(2)  '*p* but he does not believe that *p*'.

Similarly, a change of tense also results in a coherent assertion:

(3)  '*p* but I did not believe it'.

Finally, the absurdity of (1) also vanishes when the sentence is embedded in a larger context:

(4)  'Suppose that *p* but I do not believe that *p*'.

These characteristics of (1), which are not mirrored in the case of typical contradictions, seem to indicate that the problem is not a function of the truth conditions of the sentence, but rather of the performance conditions of the speech act that expresses it and, perhaps, of the mental analogue of these performance conditions for the corresponding propositional attitudes. For this reason, it has been argued that an adequate analysis of the notion of belief must be made in terms of statements, as opposed to sentences. Although an analysis in terms of sentences is all that is needed in most cases, there are certain properties of statements, such as the identity of the speaker and the recipient of his words, that cannot be defined solely in terms of the forms of words. On the other hand, an analysis in terms of sentences recommends itself due to the possibility of developing simplified logical systems which avoid the unmanageable task of defining complex performance conditions.

In *Knowledge and Belief* (1962), Hintikka tried to achieve the synthesis of both approaches. In that seminal work, Hintikka developed a multimodal logic for statements that express sentences containing the epistemic notions of knowledge and belief. Most of his analysis is made in terms of sentences, including his explanation of Moore's Paradox, but he describes the way in which the system can be expanded to handle sentences whose meaning varies according to the context of utterance and the identity of the speaker.

A conspicuous feature of Hintikka's analysis is that the logic of belief turns out to be parasitic on the logic of knowledge in the sense that the former is simply a weaker version of the latter. Although one would expect to find more than one similarity between both logics, the fact that the logic of belief is so closely modeled after the system developed for the notion of knowledge leads Hintikka to adopt a set of axioms that—I will argue—is unnecessarily strong and highly problematic. In this essay I will develop an alternative logical system for *sentences* containing the notion of belief. The system, which I will call H\*, retains the basic elements of Hintikka's system but it is based on a weaker set of axioms. I will try to show that the axioms of H\* capture in a more accurate way our logical intuitions about the notion of belief without sacrificing the possibility of providing an explanation for problematic cases such as Moore's Paradox.

**2. The System H\***

Hintikka's analysis of the notion of belief is based on the two multimodal operators $B_a$ and $C_a$, which are the formal counterparts of '*a* believes that' and 'it is compatible with everything that *a* believes that'. Each subscript *a*, *b*, *c*, ... represents a different individual. Hintikka explains the intuitive idea behind the operator $C_a$ in terms of consistency. If my beliefs are consistent, it must be possible for all of them to turn out to be true without having to give up any of them. Similarly, if something is *compatible* with my consistent beliefs, then it must be possible for this something to turn out to be the case together with everything I believe without making it necessary for me to give up any of my beliefs (1962, p. 24). In formal terms, if the set $\{B_a q_1, B_a q_2, ..., B_a q_k, C_a p\}$ is consistent, then the set $\{B_a q_1, B_a q_2, ..., B_a q_k, q_1, q_2, ..., q_k, p\}$ must also be consistent.

      This interpretation, however, is problematic. Clearly my beliefs are consistent if there is a possible state of affairs in which they are all true, but it is not obvious why it should be added that in that state of affairs I must possess those beliefs. If my beliefs are consistent, it is compatible with everything I believe that there is a state of affairs in which my beliefs are true even though I may not possess some of them. But Hintikka's interpretation of $C_a$ excludes that possibility because the set $\{B_a p, C_a \sim B_a p\}$ turns out to be inconsistent.[2] On the other hand, if my beliefs are consistent, it is also compatible with

---

[2] The reason is obvious. Suppose the set $\{B_a p, C_a \sim B_a p\}$ is consistent. Then, according to Hintikka's interpretation of $C_a$, the set $\{p, B_a p, \sim B_a p\}$ is consistent, which is absurd.

everything I believe that there is a state of affairs in which my beliefs are true and I possess those beliefs. This is the only possibility that Hintikka allows.

Since there is no obvious reason to exclude the first possibility, my interpretation of $C_a$ will be as follows: If the set $\{B_aq_1, B_aq_2, ..., B_aq_k, C_ap\}$ is consistent, then the set $\{q_1, q_2, ..., q_k, p\}$ must also be consistent. Thus, according to this interpretation, the set $\{B_ap, C_a\sim B_ap\}$ will not be inconsistent. One of the challenges in developing the system H* will be to formulate a set of axioms that capture these intuitions about the notion of belief.

The basis of H* is as follows.

*Primitive Symbols*

| | |
|---|---|
| $p_1, p_2, p_3, ..., p_k$ | [Propositional variables] |
| $\sim, B_a, B_b, ..., B_n, C_a, C_b, ..., C_n$ | [Monadic operators] |
| $\&, \vee, \supset, \equiv$ | [Dyadic operators] |

*Formation Rules*

FR1  A variable standing alone is a wff.

FR2  If $p$ is a wff, so is $\sim p$.

FR3  If $p$ and $q$ are wffs, and • is a dyadic operator, then $(p • q)$ is a wff.

FR4  If $p$ is a wff, then $B_ap$ and $C_ap$ are wffs.

*Definitions:*

[Def B]    $B_ap \equiv_{Def} \sim C_a\sim p$

[Def C]    $C_ap \equiv_{Def} \sim B_a\sim p$

*Axioms*

The theorems include all tautologies of the propositional calculus, plus the following axioms:

A1    $B_a(p \supset q) \supset (B_ap \supset B_aq)$

A2    $B_ap \supset C_ap$

A3    $B_ap \supset C_aB_ap$

*Transformation Rule*

Modus Ponens (MP)

$$\frac{p \quad p \supset q}{q}$$

Using possible world semantics, we can provide an intuitively meaningful interpretation of the system. A *model set* is a partial description of a possible world. A set μ of sentences is a model set iff it satisfies the following conditions:

(C.~)   If $p \in \mu$, then $\sim p \notin \mu$.

(C.&)   If $p \,\&\, q \in \mu$, then $p \in \mu$ and $q \in \mu$.

(C. ∨)  If $p \vee q \in \mu$, then $p \in \mu$ or $q \in \mu$ (or both).

(C.~ ~) If $\sim\sim p \in \mu$, then $p \in \mu$.

(C.~&)  If $\sim(p \,\&\, q) \in \mu$, then $\sim p \in \mu$ or $\sim q \in \mu$ (or both).

(C. ~∨) If $\sim(p \vee q) \in \mu$, then $\sim p \in \mu$ and $\sim q \in \mu$.

In order to provide a semantical interpretation of the multimodal operators $B_a$ and $C_a$, we need to make reference to more that one model set. The reason is obvious. If *p* is compatible with my beliefs, then there must be at least one state of affairs in which *p* turns out to be the case. But this state of affairs need not be identical with the one in which I believe that *p*. We will call a description of such state of affairs an *alternative* to μ with respect to *a*.

Let Ω be a set of model sets μ, μ*, μ**, ... Such set of model sets will be called a *model system*. The following conditions must be imposed on a model set μ.

(C.B)   If $B_a p \in \mu$ and if μ belongs to a model system Ω, then there is in Ω at least one alternative μ* to μ such that $p \in \mu^*$.

(C.B*)  If $B_a p \in \mu$ and if μ* is an alternative to μ in some model system Ω, then $p \in \mu^*$.

(C.C)   If $C_a p \in \mu$ and if μ belongs to a model system Ω, then there is in Ω at least one alternative μ* to μ such that $p \in \mu^*$.

(C. CB)   If $B_a p \in \mu$ and if $\mu$ belongs to a model system $\Omega$, then there is in $\Omega$ at least one alternative $\mu*$ to $\mu$ such that $B_a p \in \mu*$.

(C.B$_{\text{Def}}$)   $B_a p \in \mu$ if and only if $\sim C_a \sim p \in \mu$

(C.C$_{\text{Def}}$)   $C_a p \in \mu$ if and only if $\sim B_a \sim p \in \mu$.

There are several important differences between the system H* and the system proposed by Hintikka. Instead of axiom A3, Hintikka includes the following axiom in his system:

(5)   $B_a p \supset B_a B_a p$.

In defense of (5), he argues (p. 25) that the axiom is necessary to prove that the following sentence is a contradiction:

(6)   $B_a p \ \& \ (B_a p \supset B_a \sim B_a p)$.

Although (6) is certainly contradictory, it is also true that our system, whose set of axioms is weaker than the set of axioms in Hintikka's system, suffices to show that it is. Consider the following *reductio* of (6) in our system:

(6)   $B_a p \ \& \ (B_a p \supset B_a \sim B_a p) \in \mu$     Counterassumption
(7)   $B_a p \supset B_a \sim B_a p \in \mu$     From (6) by (C.&)
(8)   $B_a p \in \mu$     From (6) by (C.&)
(9)   $B_a \sim B_a p \in \mu$     From (7) and (8) by Modus Ponens
(10)   $B_a p \in \mu*$     From (8) by (C.CB)
(11)   $\sim B_a p \in \mu*$     From (9) by (C.B*)

(10) and (11) violate (C.~), thus reducing the counterassumption *ad absurdum*. The proof in H* shows that Hintikka's argument alone does not justify the inclusion of axiom (5) in a logic of belief. In the next section I will argue that there are independent reasons to reject (5) and to adopt the weaker set of axioms.

The conditions that Hintikka imposes on model sets also differ from the ones in our system. Instead of our condition (C.CB), Hintikka includes the following condition in his system:

(C.BB*)   If $B_a p \in \mu$ and if $\mu^*$ is an alternative to $\mu$ in some model system $\Omega$, then $B_a p \in \mu^*$.

Notice the difference between (C.BB*) and (C.CB). The former says that if I believe something in $\mu$, I believe it in every alternative to $\mu$. The latter says that if I believe something in $\mu$, there is at least one alternative to $\mu$ in which I believe it. (C.BB*) entails (C.CB), but not the converse. Intuitively, if I believe that $p$ in $\mu$, (C.CB) does not rule out the possibility of there being alternatives to $\mu$ in which I do not believe that $p$. This is not a problem, for all that is needed for my beliefs to be consistent is that there be at least one alternative model set in which they are true. If I believe that $p$, (C.B) alone guarantees that there is at least one alternative model set in which $p$ is true. (C.CB) is added to reflect our intuition that there are some model sets in which $p$ is true and I believe it, and others in which $p$ is true and I do not believe it.

The difference between (C.CB) and (C.BB*), of course, simply reflects our choice of axiom A3 instead of Hintikka's axiom (5). In our system, the sentence

(12)   $B_a p \supset C_a {\sim} B_a p$

is not a contradiction. If $B_a p \in \mu$, there may be an alternative $\mu^*$ to $\mu$ such that ${\sim}B_a p \in \mu^*$. Hence, $B_a p \supset B_a B_a p$ —Hintikka's axiom (5)—will be false in some model sets. On the other hand, according to our condition (C.CB), if $B_a p \in \mu$, then there is at least one alternative $\mu^*$ to $\mu$ such that $B_a p \in \mu^*$. Therefore, our third axiom, $B_a p \supset C_a B_a p$, is true in every model set of our system. In the remaining sections of the essay I will provide further reasons to support the claim that Hintikka's axiom (5) should not be a theorem in a logic of belief and that it should be replaced by axiom A3.

## 3. Believing That One Believes

Prima facie, Hintikka's axiom (5) seems extremely plausible. If I believe that $p$, it seems absurd to deny that I believe that I believe that $p$. But the obviousness of (5) disappears when the sentence is not in the first-person. It is not absurd to assert of someone else that he believes that $p$ but he does not believe that he believes that $p$. Consider the truth conditions for the sentence $B_a p$. The sentence is true iff $a$ is in an intentional state whose content is $p$ and whose propositional attitude is belief. $\sim B_a p$, on the other hand, is true iff $a$ is not in that intentional state. Now consider the truth conditions for the sentence $B_a B_a p$. The sentence is true iff $a$ is in an intentional state whose content is $B_a p$ and whose mental attitude is belief. In other words, $B_a B_a p$ is true iff $a$'s belief that $p$ is accompanied by a concurrent belief whose content is $B_a p$. The negation of $B_a B_a p$, on the other hand, is true iff $a$ is not in the intentional state of believing that $B_a p$.

It is perfectly possible that when someone believes that $p$, no concurrent belief occurs about that person's belief that $p$. For example, the sentence

(13)  'Mary believes that it is raining but she does not believe that she believes that it is raining'

is not absurd or self-contradictory. A virtue of our system is that it captures this important fact about the notion of belief. The sentence

(13*)  $B_a p \;\&\; \sim B_a B_a p$

is not a contradiction in H*. But notice that (13*) is simply the negation of Hintikka's axiom (5). If we accept (5) as an axiom, we would also have to claim that (13) and (13*) are contradictions (which is clearly not the case). Therefore, unless we want to misrepresent an important aspect of the analysis of the notion of belief, (5) cannot be an axiom of the system.

What, then, about the absurdity of uttering sentence (13*) when $a$ is the speaker? One option is to modify our system and include an *ad hoc* clause about the performance conditions of sentences in the first person.[3] But there is no need to do so. Notice that (13\*) is just a version of Moore's Paradox and can be treated as such. In order to examine the oddity of (13\*), we must now turn to the analysis of Moore's Paradox.

## 4. Explaining the Paradox

Before considering the problem described above, I will return to the original paradox in order to show that Hintikka's proposed solution is also valid in our weaker system. We can symbolize Moore's Paradox in the following terms:

(1\*)   $p \ \& \sim B_a p$.

The sentence is not a contradiction because

(14)   $p \supset B_a p$

is *not* a theorem in either system. Hintikka's ingenious solution to the paradox is to argue that although (1\*) is not a contradiction, the following sentence is:

(15)   $B_a(p \ \& \sim B_a p)$.

This sentence corresponds to the general presumption that the speaker believes or at least can conceivably believe what he or she says. In Hintikka's words, "the gist of Moore's Paradox may be said (somewhat elliptically) to lie in the fact that [(15)] is necessarily *unbelievable* by the speaker" (p. 67). A virtue of this approach is that nothing turns on the peculiarities of the first-person singular pronoun. (15) is contradictory no matter who *a* is.

Instead of presenting Hintikka's proof that (15) is a contradiction, I will present a similar proof in H\*, and indicate the pertinent differences. The proof is a *reductio* of the following counterassumption:

(16)   $B_a(p \ \& \sim B_a p) \in \mu$        Counterassumption
(17)   $p \ \& \sim B_a p \in \mu^*$        From (16) by (C.B\*)

---

[3] Axiom (5) makes Hintikka's system a KD4 system of epistemic logic. Rieger (2015) shows that adding the negation of a sentence that states Moore's Paradox to a KD system would be sufficient

(18)　$B_a(p \ \& \sim B_ap) \in \mu^*$　　　From (16) by (C.CB)

(19)　$\sim B_ap \in \mu^*$　　　From (17) by (C.&)

(20)　$C_a\sim p \in \mu^*$　　　From (19) by (C. $B_{Def}$)

(21)　$\sim p \in \mu^{**}$　　　From (20) by (C.C)

(22)　$p \ \& \sim B_ap \in \mu^{**}$　　　From (18) by (C.B*)

(23)　$p \in \mu^{**}$　　　From (22) by (C.&)

Here (21) and (23) contradict (C.~), thus completing the reductive argument. The main difference between our argument and Hintikka's is the justification of (18). Instead of (C.CB), which is not a condition in Hintikka's system, he uses (C.BB*). This difference is unimportant because, as Hintikka admits at one point, the proof could be done without making use of that condition.

We can now return to the version of Moore's Paradox presented in the previous section. Following Hintikka's strategy, we can prove that although (13*) is not a contradiction in H*, the following sentence is:

(24)　$B_a(B_ap \ \& \sim B_aB_ap)$.

The proof is a *reductio* of the following counterassumption:

(25)　$B_a(B_ap \ \& \sim B_aB_ap) \in \mu$　　　Counterassumption

(26)　$B_ap \ \& \sim B_aB_ap \in \mu^*$　　　From (25) by (C.B*)

(27)　$B_a(B_ap \ \& \sim B_aB_ap) \in \mu^*$　　　From (25) by (C.CB)

(28)　$\sim B_aB_ap \in \mu^*$　　　From (26) by (C.&)

(29)　$C_a\sim B_ap \in \mu^*$　　　From (28) by (C.$B_{Def}$)

(30)　$\sim B_ap \in \mu^{**}$　　　From (29) by (C. C)

(31)　$B_ap \ \& \sim B_aB_ap \in \mu^{**}$　　　From (27) by (C.B*)

(32)　$B_ap \in \mu^{**}$　　　From (31) by (C.&)

---

to block it, but that move seems equally *ad hoc*. The present proposal is based on a revision of the modal operator $C_a$ itself, which leads directly to a rejection of Axiom (5).

(30) and (32) contradict (C.~), thus completing the reductive argument. Just as in the original version of the paradox, (24) is a contradiction regardless of the identity of *a*.

This analysis of (13*) explains the problematic case in which the sentence is in the first-person and gives further plausibility to my choice of axiom A3. In fairness to Hintikka, I must admit that the advantages of my system over the one he developed in *Knowledge and Belief* depend entirely on my reinterpretation of the modal operator $C_a$. I believe, however, that this reinterpretation, and the system that can be constructed on it, give us a better picture of the logical structure of sentences containing the notion of belief.

## 5. Conclusion

Epistemic logic has come a long way since the publication of Hintikka's seminal work, and there are other ways of dissolving and explaining the paradox in the literature[4]. My purpose in this paper, however, was merely historical. I wanted to show how it is possible to modify Hintikka's original axioms and definitions to provide a system that is better attuned to our philosophical intuitions. In particular, I presented a better interpretation of the modal operator $C_a$, which leads directly to a rejection of the idea that when someone believes that *p*, there must be a concurrent belief about that person's belief that *p*. More formally, $B_a p$ & $\sim B_a B_a p$ is not a contradiction in H*. This formula is the negation of axiom (5) in Hintikka's original system. The formula is also a version of Moore's Paradox when *a* is the speaker. As I show in the last section, it can be established in H* that $B_a(B_a p$ & $\sim B_a B_a p)$ is a contradiction, so Hintikka's original explanation of the paradox remains valid in the new system. It remains to be seen how the system H* relates to Hintikka's analysis of knowledge, but that is left for future work.

---

[4] See Green and Williams (2007) for a survey.